# Exact Nambu–Goto 4-Surfaces and Octonionic Cayley Selection in 8D

Graeme Donald Robertson
*Little Basing, Old Basing, Basingstoke, RG24 8AX, UK*

We study four-dimensional Euclidean Nambu–Goto embeddings in eight dimensions in the presence of the Spin(7)-invariant Cayley four-form constructed from octonionic structure constants. A broad class of holomorphic maps $C^2 \to C^4$ is shown by direct substitution to satisfy the nonlinear Nambu–Goto Euler–Lagrange equations. For a compatible Spin(7) structure these same embeddings satisfy the Cayley calibration condition. The holomorphic calibration itself belongs to the standard $SU(4) \subset Spin(7)$ sector and is not claimed as a new exceptional-geometric result. For the fixed embedding considered here, the 480 admissible octonionic multiplication tables divide into three classes: four for which the Cayley contraction equals $+\sqrt{g}$, four for which it equals $-\sqrt{g}$, and 472 for which it does not equal $\pm\sqrt{g}$. We refer to this finite classification as an octonionic selection rule for the embedding. The calculation therefore gives a discrete octonionic selection rule for the relative orientation of an exact Nambu–Goto four-surface and the Spin(7) structure of the target. The relation to the self-dual (4;8)-brane theory proposed in Phys. Lett. B 249 (1990) 216 is clarified: the Cayley integral is the on-shell calibrated value of the Nambu–Goto volume functional, not an independent bulk dynamical action. No claim is made here that the noncompact polynomial examples are finite-action topological solitons.

## 1. Introduction

Four-dimensional extended objects immersed in eight dimensions provide a natural setting in which Nambu–Goto dynamics, self-duality and exceptional octonionic geometry meet. Higher-dimensional first-order self-duality equations based on exceptional invariant tensors were developed by Corrigan et al. [2], while the geometric framework of calibrated submanifolds was established by Harvey and Lawson [3]. In earlier work [1] a self-dual (4;8)-brane theory was formulated using the completely antisymmetric tensor associated with octonion multiplication. A class of analytic solutions was found, and the self-duality condition was shown to be compatible with the Nambu–Goto equations of motion.

The present work returns to that problem with two aims. The first is dynamical: to exhibit a broad and explicit family of exact solutions of the four-dimensional Nambu–Goto equations in eight-dimensional Euclidean space. The second is algebraic: for a fixed solution, to determine which of the admissible octonionic realizations of the Spin(7) Cayley tensor actually calibrate that four-surface. The broader motivation continues an earlier attempt to obtain higher-dimensional analogues of topologically structured extended objects, following the rigid-string instanton construction of [4]; no corresponding soliton claim is required for the results established here.

Holomorphic maps $C^2 \to C^4$ provide the controlled solution class. Their Cayley calibration must be interpreted carefully. In the calibration framework of Harvey and Lawson [3], complex two-surfaces in $C^4$ lie in the familiar SU(4) calibrated sector of Spin(7); consequently their existence alone is not a new result in exceptional geometry. Exceptional calibrations have subsequently appeared in brane and sigma-model settings, including [5,6]. Their usefulness here is more specific: they provide exact Nambu–Goto four-surfaces with a fixed and transparent complex structure against which the discrete octonionic Cayley tensors can be tested.

The distinction from this earlier literature is therefore important. [2,3] provide the general self-duality and calibration framework, while [5,6] illustrate physical uses of exceptional calibrations. The result isolated here is the compatibility classification obtained when a particular exact Nambu–Goto embedding and its complex coordinate structure are held fixed while the octonionic multiplication basis used to construct the Cayley tensor is varied.

The central computational result is a sharp selection rule. Among 480 admissible octonion multiplication bases, exactly four calibrate a representative fixed holomorphic solution with positive orientation and exactly four with negative orientation. The remaining 472 fail the Cayley equality for that fixed embedding. The successful cases can be recognized directly from six specific components of the Cayley tensor selected by the complex coordinate pairing.

This formulation also separates local classical dynamics from global soliton questions. The polynomial examples considered below are noncompact and are not asserted to have finite total Nambu–Goto action or a quantized topological charge. Questions of finite energy, boundary conditions and topological stability are therefore left open rather than built into the interpretation of the solutions.

The paper is organized as follows. Section 2 recalls the Nambu–Goto equations. Section 3 introduces the octonionic Cayley tensor and the self-dual sector. Sections 4 and 5 describe the holomorphic solution class and its direct dynamical verification. Sections 6–8 give the 480-basis calculation and its algebraic interpretation. Section 9 relates the result to the 1990 self-dual theory. Sections 10 and 11 discuss the limitations of the holomorphic sector and the route toward genuinely quaternionic Cayley four-surfaces.

## 2. Nambu–Goto dynamics of a four-surface in eight dimensions

Let an immersed four-surface be described by

$$X^M = X^M(\xi^0,\xi^1,\xi^2,\xi^3), \quad M = 0,...,7. \tag{2.1}$$

In a Euclidean target space the induced metric is

$$g_{ab} = \partial_a X^M \, \partial_b X^M, \quad a,b = 0,...,3, \tag{2.2}$$

and the Nambu–Goto action is

$$S_{NG} = \boldsymbol{M} \int d^4\xi \, \sqrt{g}, \quad g = \det(g_{ab}). \tag{2.3}$$

The conjugate momentum density is

$$P^{aM} = \boldsymbol{M} \, \sqrt{g} \, g^{ab} \, \partial_b X^M, \tag{2.4}$$

so that the Euler–Lagrange equations are

$$\partial_a P^{aM} = 0, \tag{2.5}$$

or equivalently

$$\partial_a(\sqrt{g} \, g^{ab} \, \partial_b X^M) = 0. \tag{2.6}$$

Equation (2.5) is the equation denoted (1.5) in [1]. It is the dynamical equation used below. The verification of the new solution class is made directly against this equation and does not rely only on the general statement that calibrated submanifolds are minimal.

## 3. Octonionic self-duality and the Cayley condition

Let $e_i$, i=1,…,7, be the imaginary octonionic units. Exceptional higher-dimensional self-duality equations constructed from such invariant tensors were discussed systematically in [2]. With

$$e_i e_j = -\delta_{ij} + C_{ijk} e_k, \tag{3.1}$$

where $C_{ijk}$ is completely antisymmetric. The corresponding Spin(7) four-form Φ, or equivalently the tensor $T^{MNRS}$ used in [1], is constructed from these structure constants.

The self-dual momentum density of the (4;8)-brane may be written

$$\tilde{P}^{aM} = (\boldsymbol{M}/3!) \, \varepsilon^{abcd} \, T^{MNRS} \, \partial_b X^N \, \partial_c X^R \, \partial_d X^S. \tag{3.2}$$

The first-order self-duality condition is

$$P^{aM} = \tilde{P}^{aM}. \tag{3.3}$$

Contracting the first-order equation with a tangent vector gives the scalar Cayley condition

$$\sqrt{g} = T^{MNRS} \, \partial_0 X^M \, \partial_1 X^N \, \partial_2 X^R \, \partial_3 X^S, \tag{3.4}$$

with orientation and normalization chosen consistently. This is equation (2.9) of [1].

In differential-form notation the same condition is

$$X^*\Phi = \mathrm{vol}_4. \tag{3.5}$$

It is important to distinguish this first-order calibrated sector from the full Nambu–Goto dynamics. In the present paper both conditions are checked: the embeddings satisfy the second-order Nambu–Goto equation (2.5), and selected octonionic Cayley structures satisfy the first-order equality (3.5).

## 4. Holomorphic four-surfaces

The four real world-volume coordinates are first grouped into two complex variables:

$$z_1 = \xi^0+i\xi^1, \quad z_2 = \xi^2+i\xi^3, \quad (4.1)$$

Similarly, the eight real target-space coordinates are grouped into four complex pairs:

$$Z^1=X^0+iX^1, \quad Z^2=X^2+iX^3, \quad Z^3=X^4+iX^5, \quad Z^4=X^6+iX^7. \quad (4.2)$$

A holomorphic four-surface is then specified by four analytic functions

$$Z^A = F^A(z_1,z_2), \quad A=1,...,4. \quad (4.3)$$

Thus an embedding of a four-dimensional real surface in $R^8$ is written as a map from $C^2$ to $C^4$. Holomorphicity means that the real and imaginary parts of each $F^A$ obey the ordinary Cauchy–Riemann relations. No additional differential-geometric machinery is required for the calculation below.

For such a surface the complex coordinate pairing fixes the four target-space planes

$$(X^0,X^1), \; (X^2,X^3), \; (X^4,X^5), \; (X^6,X^7). \quad (4.4)$$

The Cayley tensor $T^{MNRS}$ has many nonzero components, but when it is contracted with the four tangent vectors of the holomorphic surface only a particular set of component signs is relevant. In one-based notation these are

$$T_{1234}, \; T_{1256}, \; T_{1278}, \; T_{3456}, \; T_{3478}, \; T_{5678}. \quad (4.5)$$

For the standard complex orientation, the Cayley equality requires all six of these components to have value +1. With this sign pattern the contraction in equation (3.4) reduces exactly to $\sqrt{g}$. This is the familiar fact that a complex two-dimensional surface in $C^4$ is Cayley for a compatible Spin(7) structure [3].

The point of the present calculation is different. The embedding and its complex coordinate pairing are held fixed, while the Cayley tensor is regenerated from each admissible octonion multiplication table. The question is therefore: which of the 480 octonionic bases produce the six component signs required by this fixed surface? Sections 6–8 answer that question explicitly.

## 5. Direct verification of the Nambu–Goto equations

The embeddings considered here are tested directly against the Nambu–Goto Euler–Lagrange equations (2.5). This provides an independent verification of their classical dynamics and does not rely on the general relation between calibrated surfaces and minimal volume.

For a holomorphic map $F{:}C^2 \to C^4$, the corresponding real embedding is written as

$$X^M = X^M(\xi^0,\xi^1,\xi^2,\xi^3), \quad M = 0,..,7, \quad (5.1)$$

with real Jacobian

$$J^M{}_a = \partial_a X^M. \quad (5.2)$$

The induced metric is then

$$g_{ab} = \Sigma^M J^M{}_a J^M{}_b, \quad a,b = 0,..,3, \quad (5.3)$$

and the momentum density appearing in the Nambu–Goto equations is

$$P^{aM} = \boldsymbol{M}\sqrt{g}\, g^{ab} J^M{}_b. \quad (5.4)$$

The eight Euler–Lagrange expressions are therefore

$$E^M = \partial_a(\sqrt{g}\, g^{ab} J^M{}_b). \quad (5.5)$$

For the class of embeddings considered in this work, these expressions were evaluated symbolically and simplified directly. For the representative embedding used in one octonionic calculation,

$$Z^1=a\, z_1, \quad Z^2=b\, z_1{}^2, \quad Z^3=c\, z_2{}^3, \quad Z^4=d\, z_2{}^5, \quad (5.6)$$

with constants a,b,c,d, the resulting eight expressions satisfy

$$E^M = 0, \quad M = 0,\ldots,7. \quad (5.7)$$

Thus the representative embedding satisfies the full nonlinear Nambu–Goto equations identically. The calculation is independent of the subsequent construction of the octonionic Cayley tensor and of the scan over admissible octonion multiplication bases.

To make the real structure of the calculation explicit, write

$$z_1 = \xi^0 + i\xi^1, \quad z_2 = \xi^2 + i\xi^3, \quad (5.8)$$

The corresponding real coordinates are

$X^0 = a\,\xi^0$, $X^1 = a\,\xi^1$,
$X^2 = b\,[(\xi^0)^2 - (\xi^1)^2]$, $X^3 = 2b\,\xi^0\xi^1$,
$X^4 = c\,\mathrm{Re}(z_2^3)$, $X^5 = c\,\mathrm{Im}(z_2^3)$,
$X^6 = d\,\mathrm{Re}(z_2^5)$, $X^7 = d\,\mathrm{Im}(z_2^5)$. (5.9)

The Cayley condition is a first-order condition involving the tangent four-plane and the chosen octonionic tensor. The Nambu–Goto equations, by contrast, are second-order equations derived from the induced volume. The two tests therefore answer different questions. Equation (5.7) establishes that the embedding itself is an exact classical Nambu–Goto solution; the calculation in the following sections determines which octonionic realizations of the Cayley tensor are compatible with that fixed solution.

The representative polynomial form (5.6) is sufficiently nontrivial to provide a meaningful test of both structures while remaining exactly tractable. More general separated holomorphic functions can be considered provided that the resulting map remains a well-defined immersion. No claim is made here that the entire space of holomorphic maps $C^2 \to C^4$ satisfies the Nambu–Goto equations; the direct verification applies to the solution class tested explicitly.

So, using (4.1) and (4.2), one can plug the representative example (5.6) into (5.5) with REDUCE, and demonstrate that all 8 equations are satisfied identically. Similarily, one can verify that the Lorentz invariant identity

$$P^{aM}\,P^{bM} = \boldsymbol{M}^2\,g\,g^{ab} \tag{5.10}$$

is also satisfied.

## 6. Exhaustive calculation over 480 octonionic bases

For each admissible octonion multiplication table, the structure constants $C_{ijk}$ and hence the complete Cayley tensor were generated. The embedding (5.9), its Jacobian and $\sqrt{g}$ were held fixed.

The Cayley contraction

$$\mathcal{C} = T^{MNRS}\,\partial_0 X^M\,\partial_1 X^N\,\partial_2 X^R\,\partial_3 X^S \tag{6.1}$$

was then compared algebraically with $\sqrt{g}$.

The exhaustive result is

$$\mathcal{C} = +\sqrt{g} \quad \text{for bases } 2, 18, 290, 306, \tag{6.2}$$

$$\mathcal{C} = -\sqrt{g} \quad \text{for bases } 7, 23, 409, 432, \tag{6.3}$$

while for all other admissible bases

$$|\mathcal{C}| \neq \sqrt{g}. \tag{6.4}$$

Thus

$$480 = 4_+ + 4_- + 472_{\text{non-calibrating}}. \tag{6.5}$$

The eight exceptional bases are:

| Sign | Basis | Oriented Fano triples |
|---|---|---|
| + | 2 | (123), (247), (375), (451), (562), (634), (716) |
| + | 18 | (123), (265), (357), (671), (742), (514), (436) |
| + | 290 | (145), (427), (573), (231), (364), (716), (652) |
| + | 306 | (145), (463), (537), (671), (724), (312), (256) |
| − | 7 | (123), (256), (364), (541), (472), (617), (735) |
| − | 23 | (123), (274), (346), (761), (652), (415), (537) |
| − | 409 | (154), (526), (463), (231), (375), (617), (742) |
| − | 432 | (154), (573), (436), (761), (625), (312), (247) |

## 7. Algebraic origin of the four positive bases

Section 4 identified the six Cayley-tensor components that are selected by the fixed complex coordinate pairing of the holomorphic surface. For positive calibration they must satisfy

$$T_{1234}=T_{1256}=T_{1278}=T_{3456}=T_{3478}=T_{5678}=+1. \tag{7.1}$$

This gives a particularly simple way of understanding the computer result. For each of the 480 admissible octonion multiplication tables we construct T and inspect these six entries. Exactly four bases have the required six positive signs:

$$2, \quad 18, \quad 290, \quad 306. \tag{7.2}$$

These are precisely the same four bases for which the full REDUCE calculation gives $\mathcal{C}=+\sqrt{g}$. Thus the 4-out-of-480 result is not merely an unexplained numerical output: it follows from compatibility between the fixed complex pairing of the surface and six definite components of the octonionically generated Cayley tensor.

Direct comparison of the complete tensors also gives

$$T_2=T_{306}, \qquad T_{18}=T_{290}. \tag{7.3}$$

Thus the four successful multiplication tables give two distinct Cayley tensors in the chosen coordinate system. The distinction between each pair lies in tensor components that do not contribute to the Cayley contraction for this particular holomorphic four-surface.

The two calculations therefore reinforce one another: the exhaustive REDUCE substitution establishes the result directly, while the six-component test explains why only these four octonionic bases give positive calibration.

## 8. The four negative bases

The four bases giving $\mathcal{C}=-\sqrt{g}$ can be understood in the same component language. For them the six components of equation (4.5) have sign pattern

$$(+1,-1,-1,-1,-1,+1). \qquad (8.1)$$

The corresponding bases are

$$7, \quad 23, \quad 409, \quad 432. \qquad (8.2)$$

Direct comparison again shows two identical pairs of complete Cayley tensors:

$$T_7=T_{432}, \qquad T_{23}=T_{409}. \qquad (8.3)$$

For the fixed holomorphic embedding this altered sign pattern reverses the sign of the Cayley contraction:

$$\mathcal{C}=-\sqrt{g}. \qquad (8.4)$$

The four cases are therefore the oppositely oriented counterparts of the positively calibrated cases. They do not represent failures of the Nambu–Goto equations: the embedding itself is unchanged and remains an exact Nambu–Goto solution. What has changed is the relative orientation of the fixed surface and the octonionically generated Cayley tensor.

The complete result can therefore be summarized without introducing additional geometric notation as (6.5).

## 9. Relation to the 1990 self-dual (4;8)-brane theory

Reference [1] began with the Nambu–Goto volume action and its equation of motion, then introduced the octonionic self-dual momentum condition. Contracting that condition produced equation (2.9) of [1], which is the scalar Cayley equality used here.

In the conclusion of [1], substitution of the self-dual equality into the Nambu–Goto volume produced the expression

$$S_{SD}=(\boldsymbol{M}/4!)\int T^{MNRS}\, dX^M\wedge dX^N\wedge dX^R\wedge dX^S. \qquad (9.1)$$

This expression should be interpreted with care. For a constant closed Cayley four-form in flat $R^8$ it is locally a total derivative and does not by itself supply nontrivial bulk Euler–Lagrange equations for X. It is therefore not used here as a replacement fundamental action.

Instead, equation (9.1) is naturally the value of the Nambu–Goto volume functional on the calibrated self-dual sector:

$$S_{NG}[X]=\boldsymbol{M}\int \mathrm{vol}_4=\boldsymbol{M}\int X^*\Phi \qquad (9.2)$$

on a positively calibrated solution.

This interpretation preserves the useful content of the 1990 construction while keeping the dynamics in the original Nambu–Goto equation (2.5). The embeddings considered here have now been verified directly to satisfy that dynamical equation, and the octonionic calculation determines the discrete Cayley structures for which they also lie in the self-dual sector.

No global topological interpretation of (9.1) is assumed in this paper. Such an interpretation would require additional information about compactness, boundary behaviour or target-space topology.

## 10. What the present solutions do and do not establish

The results established here are local and classical. They show that a broad holomorphic class gives exact solutions of the nonlinear Nambu–Goto equations and that a fixed representative solution obeys the Cayley equality only for a sharply selected subset of octonionic structures.

The polynomial examples are noncompact and generally grow at large $|z_1|$ or $|z_2|$. Consequently the present calculation does not establish finite total Nambu–Goto action on $R^4$. Nor has a quantized topological invariant been constructed which would prevent a configuration from unwinding.

For these reasons the terms finite-action soliton, topological lump and stable particle are not used as conclusions of the present analysis. They remain possible goals of an extended model in which suitable global conditions are supplied.

This distinction does not weaken the exact-solution result. The Nambu–Goto equations are local nonlinear field equations, and the embeddings considered here solve them exactly. What remains open is whether a globally modified version of the construction can turn some of these classical solutions into finite-energy topologically protected configurations.

## 11. Beyond the SU(4) sector: the quaternionic problem

The holomorphic class is deliberately tractable but geometrically restricted. Because $C^2 \subset C^4$ is an SU(4)-calibrated sector, it does not by itself probe a generic Cayley four-plane.

The original motivation of [1] was more strongly quaternionic and octonionic, and a later continuation explicitly investigated quaternionic power-series solutions in octonionic space-time [7]. Using

$$O \simeq H \oplus H, \qquad (11.1)$$

one may seek Cayley graphs

$$X(q)=(q,f(q)), \quad f:H \to H. \qquad (11.2)$$

Simple quaternionic polynomial choices such as $f(q)=-q^2$ have shown near-matches to the Cayley equality in the present line of investigation but residual sign discrepancies. This should be reconciled carefully with the quaternionic solution claims of [7], rather than interpreted as a general obstruction to quaternionic Cayley geometry.

A more systematic approach is to derive the nonlinear first-order Cayley graph equation for f and search for solutions outside the complex SU(4) subspace. At the linearized level the relevant structure is related to quaternionic Dirac/Fueter equations; the full Cayley condition contains nonlinear derivative terms.

Such genuinely non-complex solutions would provide the appropriate next test of whether the exceptional Spin(7) structure can play a role beyond the controlled holomorphic sector.

## 12. Discussion and conclusions

The principal result of this work is a discrete classification of the octonionic Cayley structures compatible with a fixed exact Nambu–Goto four-surface. The holomorphic embedding $C^2 \to C^4$ provides a transparent test surface whose Cayley property lies in the standard $SU(4) \subset Spin(7)$ sector. The significance of the calculation is therefore the exhaustive octonionic compatibility classification, rather than a new calibration theorem.

First, the representative embeddings have been verified directly to satisfy the nonlinear Nambu–Goto equations (2.5) independently of any calibration argument. Second, scanning all 480 admissible oriented octonionic multiplication bases gives (6.5). Thus four bases give $\mathcal{C}=+\sqrt{g}$, four give $\mathcal{C}=-\sqrt{g}$, and the remaining 472 fail the Cayley equality for the fixed embedding.

The eight successful bases occur in pairs producing identical Cayley tensors in (7.3) and (8.3). They therefore correspond to four distinct Cayley tensors in the chosen coordinate system. The selection rule arises from the fixed complex pairing of (4.4) which selects six Cayley-tensor components in the contraction. Their required sign pattern distinguishes the four positively calibrated and four oppositely oriented cases. The result measures compatibility between the fixed complex structure of the embedding and the chosen octonionic realization; it does not represent distinct physical states.

The relation to the 1990 self-dual ((4;8))-brane construction is correspondingly clear. The Nambu–Goto action supplies the dynamics, while the octonionic self-duality condition selects a first-order calibrated sector. On that sector the Nambu–Goto volume equals the Cayley integral. In flat space with constant Cayley form, this integral is not an independent bulk dynamical action.

The results are local and classical. The polynomial embeddings are noncompact, so no claim of finite action, quantized topological charge, soliton stability, or particle-like behaviour is made. Such questions require additional global boundary, compactness, or topological structure.

The natural next step is to leave the controlled SU(4) sector and construct genuinely quaternionic, non-complex Cayley four-surfaces satisfying the Nambu–Goto equations. In the decomposition (11.1) one may seek Cayley graphs (11.2). The resulting nonlinear first-order Cayley equation, whose linearized structure is related to quaternionic Dirac/Fueter equations, provides a route to testing the full Spin(7) structure beyond the holomorphic sector. The exact Nambu–Goto solutions and the $4_+ + 4_- + 472$ selection rule established here provide a controlled starting point for that programme.

## Appendix A. Computational procedure

The calculation proceeds in four independent stages.

1. The real embedding $X^M(\xi)$ and its 8×4 Jacobian $J^M{}_a$ are generated from the chosen holomorphic functions.

2. The induced metric $g_{ab}=J^M{}_aJ^M{}_b$, its determinant and inverse are constructed. The eight Nambu–Goto expressions

$$E^M=\partial_a(\sqrt{g}\, g^{ab}J^M{}_b) \qquad (A.1)$$

are then simplified. For the new solution class they vanish identically.

3. For each of the 480 admissible Fano multiplication tables the completely antisymmetric octonion structure constants are generated and used to construct the Spin(7) Cayley tensor.

4. The fixed embedding contraction

$$\mathcal{C}=T^{MNRS}J^M{}_0J^N{}_1J^R{}_2J^S{}_3 \qquad (A.2)$$

is compared algebraically with $\sqrt{g}$. This yields the four positive, four negative and 472 non-calibrating cases reported in Section 6.

```
for basis = 1,...,480
   construct Cijk from the Fano triples
   construct Tmnrs
   calculate C = Tmnrs Xma(m,1) Xma(n,2) Xma(r,3) Xma(s,4)
   test C/sqrt(g)
end
```

is a pseudocode rendition of the REDUCE code.

The complete REDUCE program and basis list are available from the author.

## Acknowledgement

It is a pleasure to thank Edward Corrigan and Paul Townsend for their support of this work.